\documentclass[nautre,english,floatfix,twocolumn,showpacs,preprintnumbers,superscriptaddress,amsmath,amssymb]{revtex4-2}

\usepackage{graphicx}
\usepackage{dcolumn}
\usepackage{bm}
\usepackage{xcolor}
\usepackage{multirow}
\usepackage[normalem]{ulem}
\usepackage{hyperref}
\hypersetup{
    colorlinks,
    citecolor=blue,
    filecolor=blue,
    linkcolor=blue,
    urlcolor=blue
}
\hypersetup{
    colorlinks,
    citecolor=blue,
    filecolor=blue,
    linkcolor=blue,
    urlcolor=blue
}

\usepackage{color}
\definecolor{nicegreen}{RGB}{46,204,64}
\definecolor{davidblue}{RGB}{0,68,255}

\begin{document}
\title[]{Measurement of the Isolated Nuclear Two-Photon Decay in $^{72}\mathrm{Ge}$}

\author{D.~Freire-Fern\'{a}ndez}
\email[]{D.FreireFernandez@gsi.de}
\affiliation{Max-Planck-Institut f\"{u}r Kernphysik, 69117 Heidelberg, Germany}
\affiliation{Ruprecht-Karls-Universit\"{a}t Heidelberg, 69120 Heidelberg, Germany}

\author{W.~Korten}
\email[]{W.Korten@cea.fr}
\affiliation{IRFU, CEA, Universit{\'e} Paris-Saclay, Gif-sur-Yvette, 91191, France}

\author{R.~J.~Chen}
\affiliation{GSI Helmholtzzentrum f{\"u}r Schwerionenforschung GmbH, 64291 Darmstadt, Germany}
\affiliation{CAS Key Laboratory of High Precision Nuclear Spectroscopy and Center for Nuclear Matter Science, Institute of Modern Physics, Chinese Academy of Sciences, Lanzhou 730000, P. R. China}

\author{S.~Litvinov}
\affiliation{GSI Helmholtzzentrum f{\"u}r Schwerionenforschung GmbH, 64291 Darmstadt, Germany}

\author{Yu.~A.~Litvinov}
\affiliation{GSI Helmholtzzentrum f{\"u}r Schwerionenforschung GmbH, 64291 Darmstadt, Germany}

\author{M.~S.~Sanjari}
\affiliation{GSI Helmholtzzentrum f{\"u}r Schwerionenforschung GmbH, 64291 Darmstadt, Germany}
\affiliation{Aachen University of Applied Sciences, Aachen, Germany}		

\author{H.~Weick}
\affiliation{GSI Helmholtzzentrum f{\"u}r Schwerionenforschung GmbH, 64291 Darmstadt, Germany}

\author{F.~C.~Akinci}
\affiliation{Department of Physics, Istanbul University, 34134 Istanbul, Turkey}

\author{H.~M.~Albers}
\affiliation{GSI Helmholtzzentrum f{\"u}r Schwerionenforschung GmbH, 64291 Darmstadt, Germany}

\author{M.~Armstrong}
\affiliation{GSI Helmholtzzentrum f{\"u}r Schwerionenforschung GmbH, 64291 Darmstadt, Germany}
\affiliation{Institut f{\"u}r Kernphysik, Universit{\"a}t zu K{\"o}ln, 50937 Cologne, Germany}

\author{A.~Banerjee}
\affiliation{GSI Helmholtzzentrum f{\"u}r Schwerionenforschung GmbH, 64291 Darmstadt, Germany}

\author{K.~Blaum}
\affiliation{Max-Planck-Institut f\"{u}r Kernphysik, 69117 Heidelberg, Germany}

\author{C.~Brandau}
\affiliation{GSI Helmholtzzentrum f{\"u}r Schwerionenforschung GmbH, 64291 Darmstadt, Germany}

\author{B.~A.~Brown}
\affiliation{Department of Physics and Astronomy, and the Facility for Rare Isotope Beams,
Michigan State University, East Lansing, MI 48824-1321, USA}

\author{C.~G.~Bruno}
\affiliation{School of Physics and Astronomy, University of Edinburgh, Edinburgh EH9 3JZ, UK}

\author{J.~J.~Carroll}
\affiliation{DEVCOM Army Research Laboratory, Adelphi, MD 20783, USA}

\author{X.~Chen}
\affiliation{Faculty of Science and Engineering, University of Groningen, 9701 BA Groningen, The Netherlands}

\author{Ch.~J.~Chiara}
\affiliation{DEVCOM Army Research Laboratory, Adelphi, MD 20783, USA}

\author{M.~L.~Cortes}
\affiliation{Technische Universit\"{a}t Darmstadt, Department of Physics, Institute for Nuclear Physics, 64289 Darmstadt, Germany}

\author{S.~F.~Dellmann}
\affiliation{Goethe-Universit\"{a}t, 60438 Frankfurt, Germany}

\author{I.~Dillmann}
\affiliation{TRIUMF, Vancouver, BC V6T 2A3, Canada}
\affiliation{Department of Physics and Astronomy, University of Victoria, Victoria, BC V8P 5C2, Canada}

\author{D.~Dmytriiev}
\affiliation{GSI Helmholtzzentrum f{\"u}r Schwerionenforschung GmbH, 64291 Darmstadt, Germany}

\author{O.~Forstner}
\affiliation{GSI Helmholtzzentrum f{\"u}r Schwerionenforschung GmbH, 64291 Darmstadt, Germany}
\affiliation{Institute of Optics and Quantum Electronics, Friedrich Schiller University Jena, Jena, 07743, Germany}
\affiliation{Helmholtz Institute Jena, 07743 Jena, Germany}

\author{H.~Geissel}
\affiliation{GSI Helmholtzzentrum f{\"u}r Schwerionenforschung GmbH, 64291 Darmstadt, Germany}

\author{J.~Glorius}
\affiliation{GSI Helmholtzzentrum f{\"u}r Schwerionenforschung GmbH, 64291 Darmstadt, Germany}

\author{A.~G{\"o}rgen}
\affiliation{Department of Physics, University of Oslo, 0316 Oslo, Norway}

\author{M.~G\'{o}rska}
\affiliation{GSI Helmholtzzentrum f{\"u}r Schwerionenforschung GmbH, 64291 Darmstadt, Germany}

\author{C.~J.~Griffin}
\affiliation{TRIUMF, Vancouver, BC V6T 2A3, Canada}

\author{A.~Gumberidze}
\affiliation{GSI Helmholtzzentrum f{\"u}r Schwerionenforschung GmbH, 64291 Darmstadt, Germany}

\author{S.~Harayama}
\affiliation{Department of Physics, Saitama University, Saitama, 338-8570, Japan}

\author{R.~Hess}
\affiliation{GSI Helmholtzzentrum f{\"u}r Schwerionenforschung GmbH, 64291 Darmstadt, Germany}

\author{N.~Hubbard}
\affiliation{GSI Helmholtzzentrum f{\"u}r Schwerionenforschung GmbH, 64291 Darmstadt, Germany}
\affiliation{Technische Universit\"{a}t Darmstadt, Department of Physics, Institute for Nuclear Physics, 64289 Darmstadt, Germany}

\author{K.~Ide}
\affiliation{Technische Universit\"{a}t Darmstadt, Department of Physics, Institute for Nuclear Physics, 64289 Darmstadt, Germany}

\author{Ph.~R.~John}
\affiliation{Technische Universit\"{a}t Darmstadt, Department of Physics, Institute for Nuclear Physics, 64289 Darmstadt, Germany}

\author{R.~Joseph}
\affiliation{GSI Helmholtzzentrum f{\"u}r Schwerionenforschung GmbH, 64291 Darmstadt, Germany}

\author{B.~Jurado}
\affiliation{LP2I Bordeaux, CNRS-IN2P3, 33170 Gradignan, France}

\author{D.~Kalaydjieva}
\affiliation{IRFU, CEA, Universit{\'e} Paris-Saclay, Gif-sur-Yvette, 91191, France}

\author{Kanika}
\affiliation{Ruprecht-Karls-Universit\"{a}t Heidelberg, 69120 Heidelberg, Germany}
\affiliation{GSI Helmholtzzentrum f{\"u}r Schwerionenforschung GmbH, 64291 Darmstadt, Germany}

\author{F.~G.~Kondev}
\affiliation{Physics Division, Argonne National Laboratory, Lemont, IL 60439, USA}

\author{P.~Koseoglou}
\affiliation{Technische Universit\"{a}t Darmstadt, Department of Physics, Institute for Nuclear Physics, 64289 Darmstadt, Germany}

\author{G.~Kosir}
\affiliation{GSI Helmholtzzentrum f{\"u}r Schwerionenforschung GmbH, 64291 Darmstadt, Germany}

\author{Ch.~Kozhuharov}
\affiliation{GSI Helmholtzzentrum f{\"u}r Schwerionenforschung GmbH, 64291 Darmstadt, Germany}

\author{I.~Kulikov}
\affiliation{Ruprecht-Karls-Universit\"{a}t Heidelberg, 69120 Heidelberg, Germany}
\affiliation{GSI Helmholtzzentrum f{\"u}r Schwerionenforschung GmbH, 64291 Darmstadt, Germany}

\author{G.~Leckenby}
\affiliation{TRIUMF, Vancouver, BC V6T 2A3, Canada}
\affiliation{Department of Physics and Astronomy, The University of British Columbia,
Vancouver, BC V6T 1Z1, Canada}

\author{B.~Lorenz}
\affiliation{GSI Helmholtzzentrum f{\"u}r Schwerionenforschung GmbH, 64291 Darmstadt, Germany}

\author{J.~Marsh}
\affiliation{School of Physics and Astronomy, University of Edinburgh, Edinburgh EH9 3JZ, UK}

\author{A.~Mistry}
\affiliation{GSI Helmholtzzentrum f{\"u}r Schwerionenforschung GmbH, 64291 Darmstadt, Germany}
\affiliation{Technische Universit\"{a}t Darmstadt, Department of Physics, Institute for Nuclear Physics, 64289 Darmstadt, Germany}

\author{A.~Ozawa}
\affiliation{Institute of Physics, University of Tsukuba, Tsukuba, 305-8571, Ibaraki, Japan}

\author{N.~Pietralla}
\affiliation{Technische Universit\"{a}t Darmstadt, Department of Physics, Institute for Nuclear Physics, 64289 Darmstadt, Germany}

\author{Zs.~Podolyák}
\affiliation{School of Mathematics and Physics, University of Surrey, Guildford, GU2 7XH, UK}

\author{M.~Polettini}
\affiliation{Dipartimento di Fisica, Universit{\`a} degli Studi di Milano,  20133, Milan, Italy}
\affiliation{Istituto Nazionale di Fisica Nucleare, Sezione di Milano, 20133, Milan, Italy}

\author{M.~Sguazzin}
\affiliation{LP2I Bordeaux, CNRS-IN2P3, 33170 Gradignan, France}

\author{R.~S.~Sidhu}
\affiliation{Max-Planck-Institut f\"{u}r Kernphysik, 69117 Heidelberg, Germany}
\affiliation{GSI Helmholtzzentrum f{\"u}r Schwerionenforschung GmbH, 64291 Darmstadt, Germany}
\affiliation{School of Physics and Astronomy, University of Edinburgh, Edinburgh EH9 3JZ, UK}

\author{M.~Steck}
\affiliation{GSI Helmholtzzentrum f{\"u}r Schwerionenforschung GmbH, 64291 Darmstadt, Germany}

\author{Th.~St\"{o}hlker}
\affiliation{GSI Helmholtzzentrum f{\"u}r Schwerionenforschung GmbH, 64291 Darmstadt, Germany}
\affiliation{Institute of Optics and Quantum Electronics, Friedrich Schiller University Jena, Jena, 07743, Germany}
\affiliation{Helmholtz Institute Jena, 07743 Jena, Germany}

\author{J.~A.~Swartz}
\affiliation{LP2I Bordeaux, CNRS-IN2P3, 33170 Gradignan, France}

\author{J.~Vesic}
\affiliation{Jo{\v{z}}ef Stefan Institute, Ljubljana 1000, Slovenia}

\author{P.~M.~Walker}
\affiliation{School of Mathematics and Physics, University of Surrey, Guildford, GU2 7XH, UK}

\author{T.~Yamaguchi}
\affiliation{Department of Physics, Saitama University, Saitama, 338-8570, Japan}
\affiliation{Tomonaga Center for the History of the Universe, University of Tsukuba, Tsukuba, 305-8571, Ibaraki, Japan}

\author{R.	Zidarova}
\affiliation{Technische Universit\"{a}t Darmstadt, Department of Physics, Institute for Nuclear Physics, 64289 Darmstadt, Germany}
	
\date{\today}	
\begin{abstract}
The nuclear two-photon or double-gamma ($2\gamma$) decay is a second-order electromagnetic process whereby a nucleus in an excited state emits two gamma rays simultaneously. 
To be able to directly measure the $2\gamma$ decay rate in the low-energy regime below the electron-positron pair-creation threshold, 
we combined the isochronous mode of a storage ring with Schottky resonant cavities.
The newly developed technique can be applied to isomers with excitation energies down to $\sim100$\,keV and half-lives as short as $\sim10$\,ms.
The half-life for the $2\gamma$ decay of the first-excited $0^+$ state in bare $^{72}\mathrm{Ge}$ ions was determined to be $23.9\left(6\right)$\,ms, which strongly deviates from expectations.
\end{abstract}
\maketitle
The nuclear two-photon decay or double-gamma ($2\gamma$) involves the decay of an excited nucleus through the simultaneous emission of two $\gamma$ rays via the virtual excitation of intermediate states. The partial half-life of this decay gives access to observables such as the (transitional) electromagnetic polarizability and susceptibility, which are important ingredients in constraining parameters of the nuclear equation of state\,\cite{EOS}, determining the neutron skin thickness\,\cite{NeutronSkin}, and constraining the nuclear matrix elements of the neutrinoless double-beta decay\,\cite{ROMEO}.

This second-order quantum-mechanical process, initially formulated for the case of atomic transitions by M.~G\"{o}ppert-Mayer\,\cite{Goeppert-1929, Goeppert-Mayer-1931}, was later extended to nuclear transitions. 
The early theoretical description utilizing second-order perturbation theory\,\cite{Grechukhin-1963, Eichler-1974} was completed by Friar {\it et al.}\,\cite{Friar-1974, Friar-1975} and later generalised by considering not only dipole but also higher multipolarities by Kramp {\it et al.}\,\cite{Kramp-1987}, who derived the total $2\gamma$ decay probability for $0^{+}\rightarrow0^{+}$ ($E0$) transitions (see Eq.\,(A.42) in \cite{Kramp-1987}):
\begin{equation}
  W_{\gamma\gamma}=\frac{\omega_0^{7}}{105\pi}\left[\alpha_{E1}^{2}+\chi_{M1}^{2}+\omega_0^{4}\frac{\alpha_{E2}^{2}}{4752}\right] = \frac{\omega_0^{7}}{105\pi}M_{\gamma\gamma}^{2},
  \label{eq:2gd}
\end{equation}
where $\omega_0$ denotes the energy difference between the initial and final state, while 
$\alpha$ and $\chi$ denote the electric transition polarizability and the magnetic transition susceptibility, respectively. The sum of terms within the brackets is equivalent to the squared magnitude of the cumulative nuclear matrix element, denoted as $M_{\gamma\gamma}^{2}$.
These observables describe the difference of the electric polarizabilities and magnetic susceptibilties between the two $0^{+}$ states and are complementary to the standard nuclear polarizabilities and susceptibilities, which describe the response of the nucleus to a perturbation by electromagnetic fields, and are related to changes  of the nuclear charge distribution and currents inside the nucleus.

All previous experiments conducted to date have employed $\gamma$-ray spectroscopy in order to investigate the two-photon decay, as demonstrated in the studies referenced in \cite{PhysRevLett.7.170, PhysRev.135.B294, PhysRevC.2.462, PhysRevC.7.322, PhysRevC.1.1025}. However, the main challenge lies in distinguishing the relatively small signal of the two simultaneously emitted photons from other (direct or indirect) photon sources, such as single-photon decay, internal pair creation (IPC) or internal-conversion (IC) electrons, due to the continuous energy spectrum associated with the two-photon emission. Therefore, ideally, the search for nuclear $2\gamma$ decays is conducted in even-even (e-e) nuclei with a first excited $0^{+}$ state, since the emission of a single $\gamma$-ray is forbidden. 
In fact, the only cases where the $2\gamma$ decay of a $0^{+}\rightarrow0^{+}$ transition was successfully observed using $\gamma$-ray spectroscopy are $^{16}\mathrm{O}$, $^{40}\mathrm{Ca}$, and $^{90}\mathrm{Zr}$\,\cite{Schirmer-1984, Kramp-1987}. In these cases, the excited $0^{+}$ states are located at high excitation energies and the observed branching ratios for the $2\gamma$ decay are of the order of $10^{-4}$.
The most surprising result is that the contribution from two subsequent $M1$ ($2M1$) transitions and $2E1$ transitions are of a similar strength, while naively a dominance of the $2E1$ decay would be expected.
This has been explained by a strong cancellation effect in the electric-dipole  transition
polarizability in these doubly-magic nuclei.
This cancellation effect is related to the different structure of the two $0^{+}$ states, i.e. different contributions from $0p-0h$ and $np-nh$ excitations across the closed shells \cite{Brown23}. 

Excited $0^+$ states at low energy are a very important evidence for nuclear shape coexistence\,\cite{Heyde-2000}, however their identification is challenging since it usually requires observation of conversion electrons. This is experimentally much more difficult than detecting $\gamma$-rays due to the strong background from atomic electrons, abundantly produced in nuclear reactions, or from $\beta$-decay.
Such states are only known in a handful of nuclei across the whole nuclear chart\,\cite{Kibedi-2022}, 
among which only $^{72}\mathrm{Ge}$ and $^{98}\mathrm{Mo}$ are stable. 
Low-lying $0^+$ states, i.e. at energies below the IPC threshold ($1.022$\,MeV), have been found in unstable nuclei, often located far from the valley of stability, and thus requiring the use of radioactive ion beams. 
One of the most notable cases is $^{186}\mathrm{Pb}$,
where two excited $0^{+}$ states below $1$\,MeV have been discovered and interpreted as evidence for a unique triple shape coexistence\,\cite{Andreyev-2000}.
Another notable example is the N=Z nucleus $^{72}\mathrm{Kr}$ where the existence of an excited $0^+$ state has been speculated for several decades, before it was finally seen in conversion-electron spectroscopy after a fragmentation reaction\,\cite{Bouchez-2003}. In view of the experimental difficulties, it is possible, or maybe even likely, that many more unstable nuclei with a low-lying first excited $0^{+}$ state exist, but have so far escaped observation. Besides their importance for a better understanding of the nuclear structure, $0^{+}$ isomers in N$\approx$Z nuclei, like $^{72}\mathrm{Kr}$, may also play a vital role in the nucleosynthesis since these nuclei present waiting points in the rapid-proton capture ({\it rp}) process \cite{Schatz-1998, Sun-2005} and such isomers may influence the proton-capture rates.


Low-lying $0^{+}$ states in medium-mass e-e nuclei have typical lifetimes in the order of a few ten to hundred nanoseconds\,\cite{Garg-2023} because the $E0$ decay in these nuclei proceeds entirely via IC and therefore is a relatively slow process\,\cite{Kibedi-2022}. However, the $2\gamma$ decay width varies strongly with the excitation energy, see Eq.\,(\ref{eq:2gd}), leading to extremely small branching ratios ($<10^{-6}$) for $\omega_0<1$\,MeV. Until now, direct searches for the $2\gamma$ emission from lower-energy $0^{+}$ states were unsuccessful, reporting only upper limits\,\cite{Henderson-2014}. A $2\gamma$ decay at energies below $1$\,MeV was exclusively observed from the $11/2^{-}$ isomer in $^{137}\mathrm{Ba}$ using the fast-timing method \cite{Walz-2015,Soederstroem-2020}. Here, the single-photon decay is strongly hindered due to its highly unfavorable multipolarity ($M4/E5$).

Alternatively, if all bound electrons are removed the IC is disabled\,\cite{Litvinov-2003} and therefore $0^+$ states can only decay by $2\gamma$ emission to the ground state or by particle emission ($\alpha$- or $\beta$-decay) in unstable nuclides. 

In this Letter we report the first direct measurement of the $2\gamma$ decay of the first excited $0^+$ state in stored, fully-ionized $^{72}\mathrm{Ge}^{32+}$ nuclei. This isomer, with an excitation energy of $691.43\left(4\right)$\,keV\,\cite{ENSDF}, possesses a half-life of $444.2\left(8\right)$\,ns in neutral atoms \cite{BRAUN1984}. However, when it is fully ionised, the partial half-life for this isolated decay can be estimated to extend to several hundred milliseconds, using the average value of the previously determined $M_{\gamma\gamma}$ matrix elements\,\cite{Kramp-1987, Schirmer-1984}. 
By combining the isochronous mode of a storage ring with non-destructive single-ion-sensitive resonant Schottky detectors , the new experimental technique termed combined Schottky plus Isochronous Mass Spectrometry (S+IMS), we were able to resolve the isomeric state and measure the time evolution of the number of observed isomers with a resolution of the order of ms.
The thereby developed method is a sensitive approach to search for unknown excited $0^+$ states in exotic nuclei and for the measurement of their $2\gamma$ decays. 

The experiment was conducted at the Gesellschaft f\"{u}r Schwerionenforschung (GSI) accelerator facility, using a primary $^{78}\mathrm{Kr}$ beam that was accelerated to a kinetic energy of precisely $441$\,MeV/nucleon (see below) using the heavy-ion synchrotron SIS-18. After fast extraction the beam was impinged on a $10$\,mm thick $^9\mathrm{Be}$ production target placed in the transfer beamline towards the Experimental Storage Ring (ESR). 
Few-nucleon removal reactions at relativistic energies are known to produce low-lying isomeric states with relatively high probability of 
up to $10$\,\%. 
The $^{72}\mathrm{Ge}$ fragments emerged from the target with a mean energy of $367.9$\,MeV/nucleon.
At this relativistic energy the $^{72}\mathrm{Ge}$ ions were fully ionized and were transported and injected into the ESR.
No additional beam purification was necessary, allowing the storage of a large number of fragments within the acceptance of the ESR.

The stored ions revolved in the ESR with frequencies of about $2$\,MHz. The revolution frequency ($f$), can be related to the mass-over-charge ($m/q$) ratio of the ions through the
equation for storage ring mass spectrometry \cite{Franzke-2008}:
\begin{equation}
\frac{\Delta{f}}{f}=-\frac{1}{\gamma_t^2}\frac{\Delta(m/q)}{m/q}+\frac{\Delta v}{v}\Bigg(1-\frac{\gamma^2}{\gamma_t^2}\Bigg)\,,
\label{eq:basic}
\end{equation}
where $v$ and $\gamma$ are the velocity and the Lorentz factor of the ions, respectively. The machine parameter $\gamma_t$ is related to the relative change of the orbit length, $C$, caused by a relative change of magnetic rigidity, $B\rho=mv\gamma/q$\,\cite{Steck-2020}.
In this experiment a setting of $\gamma_t\approx1.396$ was used. 

In order to measure short-lived states ($T_{1/2}$ $< 1$\,s), the ESR was tuned in the isochronous ion-optical mode, where the energy of the $^{72}\mathrm{Ge}$ ions corresponds to the condition $\gamma\approx\gamma_t$.
Hence, the term containing the velocity spread $\Delta v/v$ in Eq.\,(\ref{eq:basic}) vanishes and the revolution frequency is a measure of the mass-to-charge ratio, $m/q$, enabling isochronous mass spectrometry (IMS)\,\cite{iso1}.
Conventionally, the revolution frequencies in the IMS are measured by foil-based time-of-flight detectors\,\cite{Trotscher-1992,Mei-2010}, which, however, lead to a rapid loss of stored particles. To enable lifetime measurements of exotic short-lived nuclides, non-destructive highly-sensitive cavity-based resonant Schottky detectors have been continuously developed at GSI in the course of the last decade\,\cite{NOLDEN2011, Sanjari_2013,Sanjari-410}. These cavities react on the relativistic highly-charged ions repeatedly passing through.
The latest edition, resonant at $410$\,MHz\,\cite{Sanjari-410}, at about the $212^{\rm th}$ harmonic of the revolution frequency, has been employed for the first time in this experiment.
Furthermore, a second, less-sensitive detector, resonant at $245$\,MHz\,\cite{Sanjari_2013} was also used. 
The power outputs from both detectors were amplified and fed into commercial real-time spectrum analyzers (RSA), which were set to digitize the power signal in a frequency span of $\pm\,20$\,kHz around the chosen central frequency with a sampling rate of $50$\,kHz. 
The central frequency was set on the revolution frequency of $^{72\mathrm{g}}\mathrm{Ge}^{32+}$. 
The data was processed with the code \textit{iqtools}\,\cite{iqtools} and the particle identification was realised with the code \textit{rionid}\,\cite{rionid}.

\begin{figure}
	\includegraphics[width=0.48\textwidth]{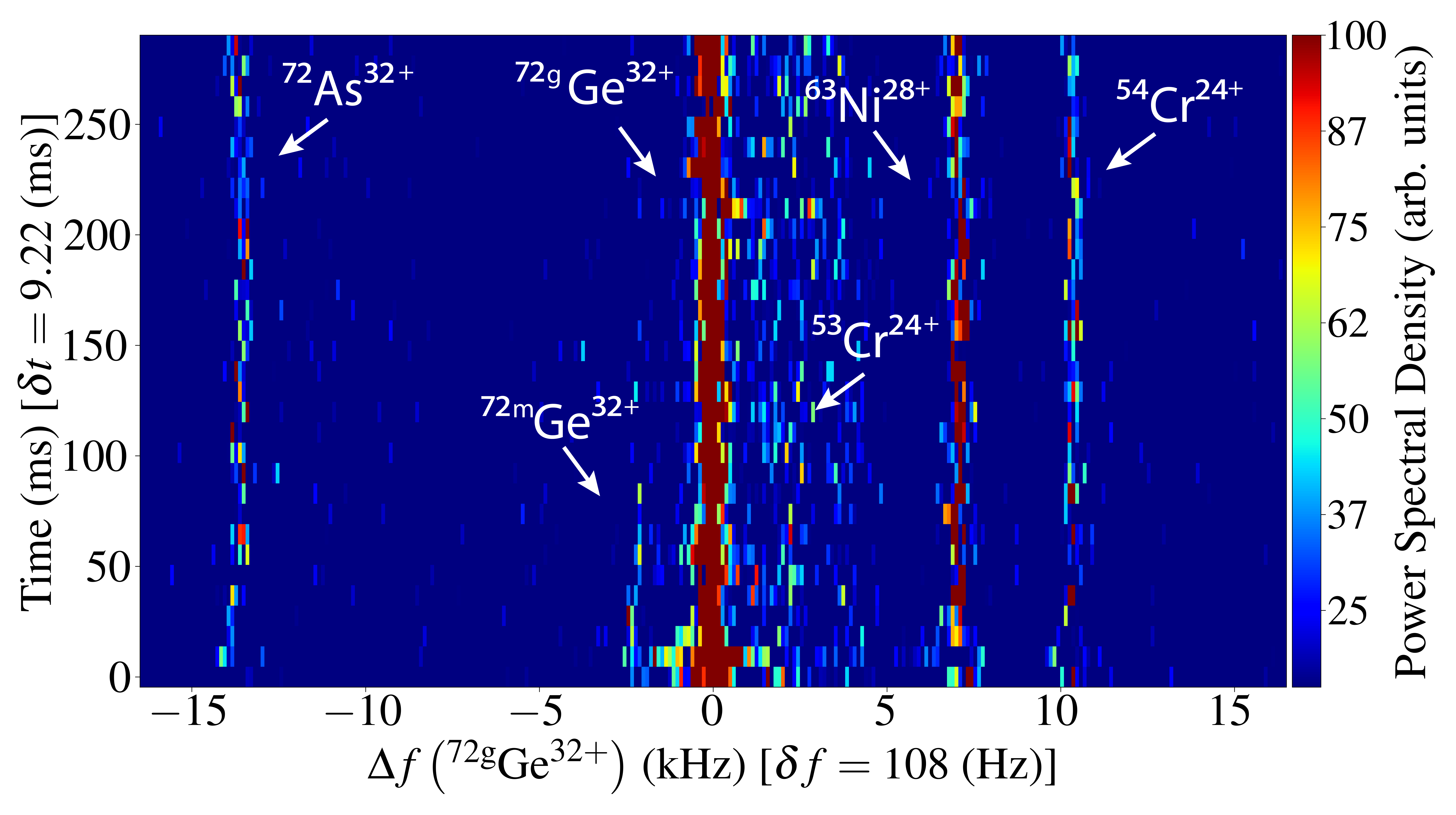}
	\caption{Frequency spectrogram of a single injection centered on $^{72\mathrm{g}}\mathrm{Ge}^{32+}$ from $0$ to $300$\,milliseconds. The power spectral density of each ion species is linearly proportional to their ion number.}
	\label{fig:single}
\end{figure}

\begin{figure}
	\includegraphics[width=0.48\textwidth]{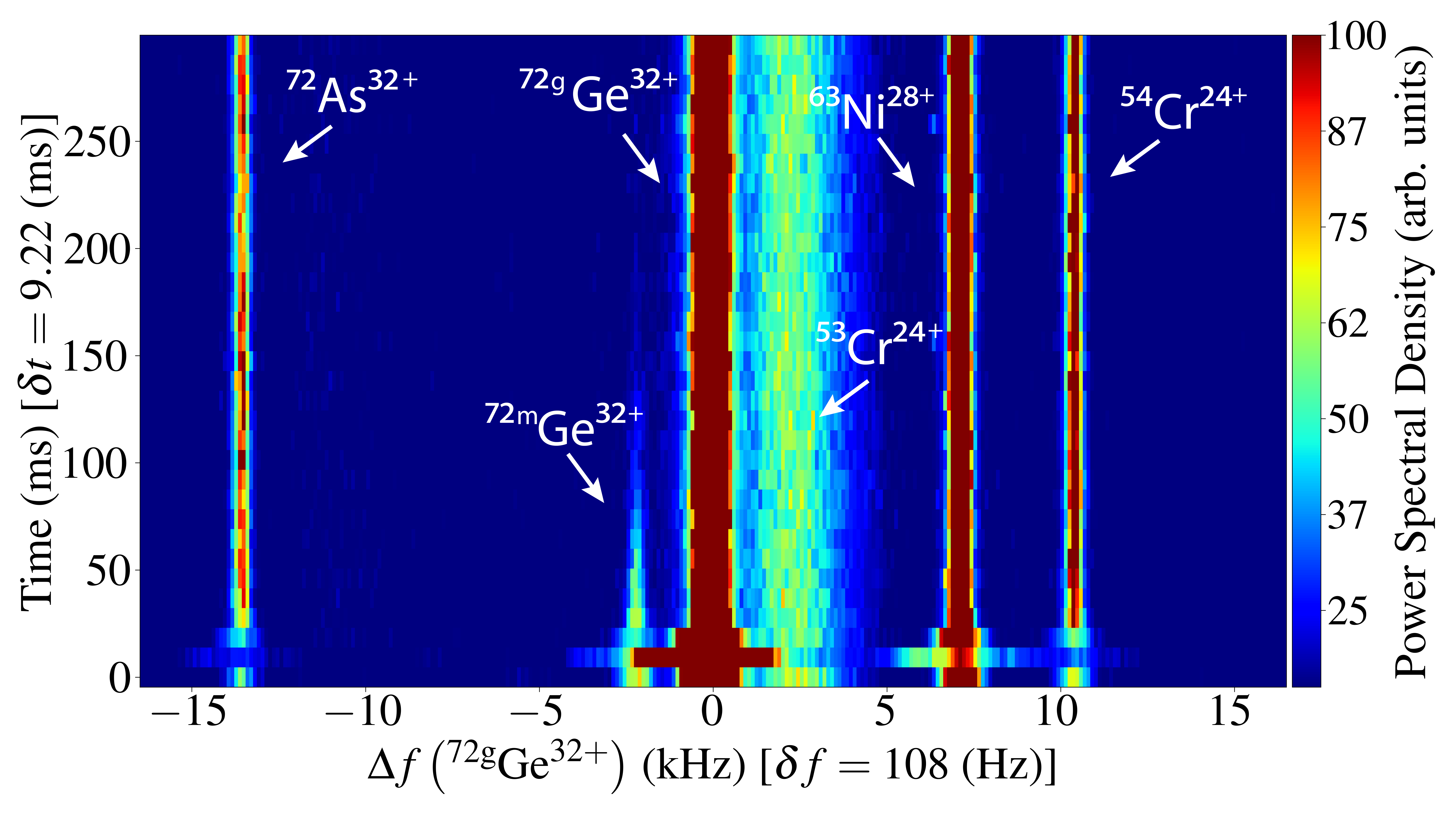}
	\caption{Frequency spectrogram of the sum of 102 single injections such as the one in Fig. \ref{fig:single}.}
	\label{fig:combined}
\end{figure}

Combinining the isochronous mode of a storage ring with the single-ion-sensitive Schottky detectors has been realised in the past\, \cite{GSIreport, CSReIso}, but was never applied to short-lived isotopes. Two high-resolution settings were achieved by using scrapers to tailor the momentum distribution of the stored ions of interest and finally, allowing us to resolve the doublet composed of the isomer $^{72\mathrm{m}}\mathrm{Ge}^{32+}$ and the ground state $^{72\mathrm{g}}\mathrm{Ge}^{32+}$ (see Figs.\,\ref{fig:single},\ref{fig:combined}).
The measurements on $^{72}\mathrm{Ge}$ were performed in two separate runs, denoted as $i=(1,2)$, which differed slightly in the achieved resolving power.
The collected data contained ($1$) $102$ injections ($\sim 30$\,min of data taking) and ($2$) $2459$ injections ($\sim 9$\,h of data taking). 

Figure\,\ref{fig:single} shows a frequency spectrogram of one measurement (single injection), where the time after injection is plotted against the revolution frequency.  
The traces corresponding to $^{72}\mathrm{As}^{32+}$, $^{72\mathrm{m}}\mathrm{Ge}^{32+}$, $^{72\mathrm{g}}\mathrm{Ge}^{32+}$, $^{53}\mathrm{Cr}^{24+}$, $^{63}\mathrm{Ni}^{28+}$ and $^{54}\mathrm{Cr}^{24+}$ are indicated.
The slight distortion at the beginning of the measurement is due to kicker chamber eddy currents caused by switching off the injection magnets. Since it is the same for the traces of all stored ion species, it can be corrected for, except for the second time bin in which the shift is too fast to resolve it.
We note that the observation of this distortion became possible thanks to the high temporal resolution of the new technique.

The measured frequencies and precisely known mass of $^{72}\mathrm{Ge}^{32+}$\,\cite{AME-2020}, allowed us to independently determine the excitation energy of the isomeric state.
Data from both RSAs and for both settings ($1$,$2$) were analyzed separately.
The results are listed in Table\,\ref{tab:isomer_energy}.
All obtained $\omega_0$ values are in good agreement with the tabulated excitation energy \cite{ENSDF}.

\begin{table}
  \centering
  \begin{tabular}{|c|c|c|c|c|} 
    \hline 
    Detector & $i$ & $\Delta f$ /\,Hz & $\omega_0$ /\,keV & \multicolumn{1}{c|}{$\overline{\omega_{0}}$ /\,keV}\\ \hline
    \hline
    \multirow{2}{*}{SD$_{410}$} & $1$ & $2162\left(8\right)$ & $694.4\left(31\right)$ & \multirow{2}{*}{$692.8\left(19\right)$} \\ \cline{2-4}
    & $2$ & $2157\left(4\right)$ & $691.8\left(24\right)$ & \\ \hline
    \multirow{2}{*}{SD$_{245}$} & $1$ & $1301\left(7\right)$ & $699.7\left(43\right)$ & \multirow{2}{*}{$695.0\left(32\right)$} \\ \cline{2-4}
    & $2$ & $1290\left(4\right)$ & $692.8\left(29\right)$ & \\ \hline
  \end{tabular}
  \caption{Measured frequency difference $\Delta f$ between the isomeric and ground state and the deduced excitation energy $\omega_0$ of the isomer for the two different data subsets ($i$) in each Schottky detector. The last column corresponds to the weighted average of the values obtained with each detector.}
  \label{tab:isomer_energy}
\end{table}

The trace of the isomer $^{72\mathrm{m}}\mathrm{Ge}$ in Fig.\,\ref{fig:single} corresponds to a single particle. 
Injections containing up to three $^{72\mathrm{m}}\mathrm{Ge}$ particles have been recorded.
Individual decay times can be determined analogously to the single-particle decay spectroscopy method utilized in studies of electron capture decays in the ESR\,\cite{Litvinov-2008a, Kienle-2013, Ozturk-2019}.
Alternatively, the spectra of individual injections can be summed up. 
The integrated noise power of every peak in the summed spectrum is then directly proportional to the corresponding number of stored particles.
The validity of this analysis approach has been thoroughly tested by dedicated simulations and 
cross-checked by extracting individual decay times, the details of which will be published elsewhere.

\begin{figure}
    \centering
    \includegraphics[width=0.48\textwidth]{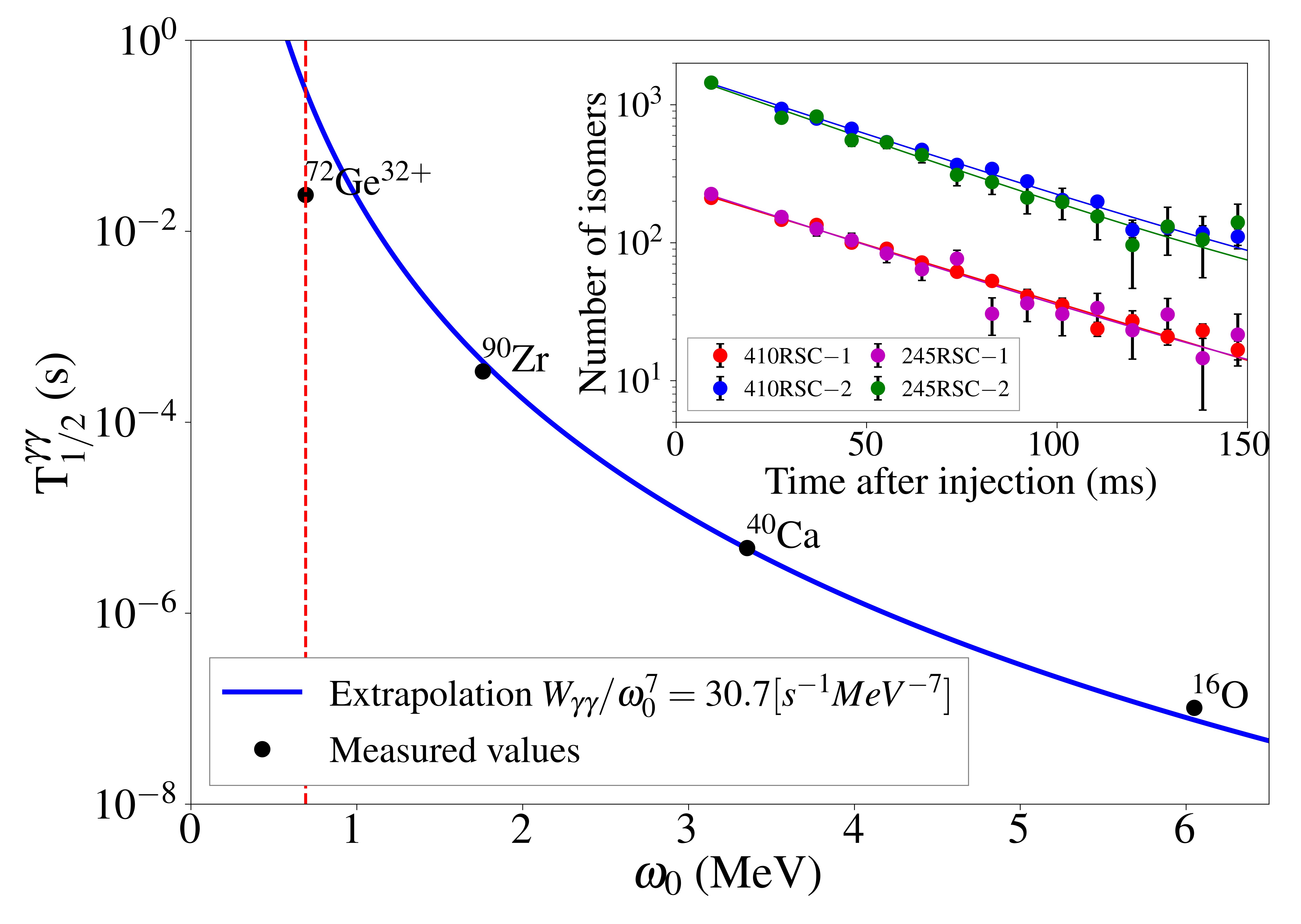}
    \caption{Measured nuclear two-photon decay half-lives, taken from \cite{Schirmer-1984, Kramp-1987} and the present work, as a function of their excitation energy. The solid line corresponds to the curve obtained by considering the ratio of $W_{\gamma\gamma}$ and $\omega_{0}^{7}$ constant, with the constant being the average value of the sum of squares of the nuclear polarizabilities in Eq.~(\ref{eq:2gd}). The vertical red dotted line is placed at the excitation energy of the isomeric state of $^{72}\mathrm{Ge}$. The insert figure shows the evolution of the number of isomers in time for each setting in each detector used.}
    \label{fig:2g-all}
\end{figure}

To determine the partial half-life of the isomer, all spectra measured by both RSAs were adjusted in frequency by setting the center of the $^{72\mathrm{g}}\mathrm{Ge}$ peaks to $0$\,Hz.
Afterwards, the spectra were summed up separately for each RSA and data set. 
The resultant combined spectra of the $410$\,MHz detector during setting $1$ is shown in Fig.\,\ref{fig:combined}.
The decrease in intensity of the $^{72\mathrm{m}}\mathrm{Ge}$ trace with time is evident.
The peak areas for each time bin were extracted and fitted with an exponential function, see the insert in Fig.\,\ref{fig:2g-all}.
The intensities of the stable species remained constant thus indicating no considerable ion losses during the measurement period.
The $245$\,MHz detector has a smaller quality factor than the $410$\,MHz one, which translates into a poorer signal-to-noise ratio and enlargement of scatter earlier in time due to background fluctuations.
The derived lifetimes are tabulated in Table\,\ref{tab:half-life}. All values agree within the uncertainties. The average measured half-life for the $2\gamma$ in the rest frame is $T_{1/2}^{\mathrm{rest}} = 23.9\left(6\right)$\,ms determined by using the Lorentz factors of the isomeric states for each setting, $\gamma_1 = 1.3954\left(1\right)$ and $\gamma_2 = 1.3959\left(1\right)$. 

\begin{table}
  \centering
  \begin{tabular}{|c|c|c|c|c|} 
    \hline 
    Detector & $i$ & $\tau^{\mathrm{lab}}$ /\,ms & $T_{1/2}^{\mathrm{rest}}$ /\,ms & \multicolumn{1}{c|}{$\overline{T}_{1/2}^{\mathrm{rest}}$ /\,ms} \\ \hline
    \hline
    \multirow{2}{*}{SD$_{410}$} & $1$ & $51.0\left(35\right)$ & $25.4\left(17\right)$ & \multirow{2}{*}{$23.9\left(6\right)$} \\ \cline{2-4}
    & $2$ & $47.7\left(13\right)$ & $23.7\left(6\right)$ &  \\ \hline
    \multirow{2}{*}{SD$_{245}$} & $1$ & $48.1\left(41\right)$ & $23.9\left(20\right)$ & \multirow{2}{*}{$22.7\left(11\right)$} \\ \cline{2-4}
    & $2$ & $44.5\left(28\right)$ & $22.1\left(14\right)$ & \\ \hline
  \end{tabular}
  \caption{Measured lifetimes in the laboratory frame and half-lives in the rest frame for the different data subsets ($i$) in each Schottky detector. The last column corresponds to the weighted average of the values obtained with each detector. The insert of Fig.~\ref{fig:2g-all} shows the experimental data from which the decay constants have been obtained.}
  \label{tab:half-life}
\end{table}

From the parameters of the exponential fits, we can compute the initial number of isomers at $t=0$\,s. Together with the number of ions in the (stable) ground state we obtain an isomeric ratio of $3.4\left(2\right)\,\%$.

In Fig.~\ref{fig:2g-all} the newly obtained partial half-life for the $2\gamma$ decay in $^{72}\mathrm{Ge}$ is plotted together with the previous results on other nuclei for $0^{+}\rightarrow0^{+}$ $2\gamma$ transitions. The solid line is an extrapolation of these previously obtained results for higher excitation energies and thus shorter half-lives to the low-energy region using the scaling suggested by Eq.\,(\ref{eq:2gd}). For the case of the first excited state in $^{72}\mathrm{Ge}$, the measured partial half-life is approximately ten times shorter than suggested by the scaling law. This implies that the sum of all contributions in Eq.\,(\ref{eq:2gd}) ($\mathrm{M}_{\gamma\gamma}=70(2)\times\,10^{-3}\,\mathrm{fm}^3$) is larger than the ones obtained in previous experiments. However, without measuring the angular correlation of the $\gamma$-rays, as could be done in Refs.\,\cite{Schirmer-1984, Kramp-1987}, it is not possible to determine $\alpha_{E1}^2$ and $\chi_{M1}^2$ individually.
We therefore have to rely on shell-model calculations  to estimate the $M1$ contribution. The transitional magnetic-dipole susceptibility consists of a paramagnetic and a diamagnetic term \cite{Friar-1974,Friar-1975, Kramp-1987}:

\begin{flalign}
   & \chi_{M1}^2= \chi_{p}^2+\chi_{d}^2\,, &
 \label{eq:chim1}
\end{flalign}

\begin{flalign}
    & \chi_{p}= -\frac{8\pi}{9} 
    \sum_n{
    \frac{\langle0_1^+\| M(M1) \| 1_n^+\rangle \, \langle 1_n^+\| M(M1) \|0_2^+ \rangle}
    {E\left(1_n^+\right)-E\left(0_1^+\right)-\frac{1}{2} 
    \left[E\left(0_2^+\right) - E\left(0_1^+\right)\right] }}\,, &
    \label{eq:chip}
\end{flalign}

\begin{flalign}
    & \chi_{d}= -\; \frac{e^2}{6m}
    \langle0_1^+\| r^2 \| 0_2^+\rangle\,. &
    \label{eq:chid}
\end{flalign}

Shell-model calculations using the $jj44$ model space and the JUN45 Hamiltonian \cite{jun45}, and an effective $M1$ operator \cite{jun45}, were performed to determine the paramagnetic contribution. 
In neighbouring $^{74}$Ge the measured $M1$ strength distribution  up to $5$\,MeV is in good agreement with these calculations \cite{joh23}.
Adding up the contributions from all theoretically expected $1^+$ states up to an excitation energy of $7.5$\,MeV results in a contribution from the magnetic-dipole transition susceptibility of $\mid\chi_{p}\mid\,\approx\,3.2\,\times\,10^{-3}$\,fm$^{3}$.
There is a strong cancellation in the terms in Eq.\,(\ref{eq:chip}) as a function of $E_{i}$; if one adds just the magnitudes, the result is about four times larger. The $jj44$ model space does not include $M1$ strength coming from the $0f_{7/2}$ to $0f_{5/2}$ contribution.
The diamagnetic contribution can be estimated from the known $E0$ matrix element \cite{Kibedi-2022} as $\mid \chi _{d}\mid\,\approx\,0.58\times10^{-3}$\,$\mathrm{fm}^3$. Therefore, the total transitional magnetic-dipole susceptibility is $\mid \chi_{M1}\mid\,\approx\,3.3\,\times\,10^{-3}$\,fm$^{3}$, which is small compared to the measured value of $M_{\gamma\gamma}$ and not too different to the values obtained by Kramp {\it et al.} \cite{Kramp-1987} in doubly-magic nuclei.

The electric transition polarizabilities for $0_2^{+}\rightarrow0_1^{+}$ transitions are defined by Eq.\,(A.24) in \cite{Kramp-1987} as:
\begin{widetext}
    \begin{equation}
    \alpha_{EL} = \frac{8\pi}{(2L + 1)^2}\sum_{n} \frac{\left\langle 0_1^+ \middle\| {i}^{L} M(EL) \middle\| 1_{n}^{{(-1)}^L} \right\rangle \left\langle 1_{n}^{{(-1)}^L} \middle\| {i}^{L} M(EL) \middle\| 0_2^+ \right\rangle}{E\left(1_{n}^{{(-1)}^L}\right)-E\left(0_1^+\right)-\frac{1}{2} 
    \left[E\left(0_2^+\right) - E\left(0_1^+\right)\right]}.
    \label{eq:alphaEL}
    \end{equation}
\end{widetext}

For the case of low-lying $0^{+}$ states, the electric-quadrupole transition polarizability $\alpha_{E2}$ may also become important. By using the reduced $E2$ matrix elements from both $0^{+}$ states to the two lowest-lying $2^{+}$ states of $^{72}\mathrm{Ge}$ \cite{AYANGEAKAA2016}, which usually exhaust more than $90$\,\% of the $E2$ strength, $\alpha_{E2}$ can be estimated via Eq.\,(\ref{eq:alphaEL}) to be $\mid \alpha_{E2}\mid\,\approx\,4749$\,fm$^{5}$. Therefore, the nuclear matrix element associated with $2E2$ transitions
amounts to:
\begin{equation}
    M_{_{E2}} = \sqrt{\omega_0^{4}\frac{\alpha_{E2}^{2}}{4752}}\,\approx\,0.8\,\times\,10^{-3}\,\mathrm{fm}^{3}.
\end{equation}

Obtaining a theoretical estimate for the electric-dipole transition polarizability via shell-model calculations requires including contributions from orbital transitions related to the giant dipole resonance region that lie outside of the $jj44$ model space. Other theoretical approaches, such as the quasiparticle random-phase approximation, are also not applicable here since the two $0^+$ states in $^{72}\mathrm{Ge}$ are strongly mixed\,\cite{Ayangeakaa-2016}. More complete set of calculations for the double-gamma matrix elements remains to be carried out. However, in view of the theoretically expected small contribution from the magnetic-dipole susceptibility and from the electric-quadrupole polarizability, it can be concluded that the electric-dipole transition polarizability is largely dominating the observed increase in transition strength.
This finding is consistent with the presumed cancellation effect of the electric-dipole transition polarizability in the case of the doubly-magic nuclei, which should not be as pronounced in this mid-shell nucleus $^{72}\mathrm{Ge}$. 

In summary, we reported the results of the combined Schottky + Isochronous Mass Spectrometry (S+IMS) applied to the direct determination of the excitation energy and partial 
half-life of an isomer in the millisecond regime with high precision. This represents a dramatic extension of the storage-ring based non-destructive lifetime spectroscopy to shorter-lived species as compared to previous experiments with electron-cooled stored beams\,\cite{Litvinov-2011}. A mass resolving power of $9.1\times10^5$ has been achieved, which allows us to fully resolve low-lying ($\sim100$\,keV) isomers. For the $2\gamma$ decay of the $0^+$ isomer in $^{72}\mathrm{Ge}$, a partial half-life of $T_{1/2}^{\mathrm{rest}} = 23.9\left(6\right)$\,ms and an excitation energy of $\omega_0=692.8\left(19\right)$\,keV have been determined. 
The obtained partial half-life is a factor $\sim10$ shorter than expected from the extrapolation of previous results based on 
doubly-magic $^{16}\mathrm{O}$, $^{40}\mathrm{Ca}$ and semi-magic $^{90}\mathrm{Zr}$, indicating that the cancellation effect observed in those nuclides appears not to be present. 
This technique will be further developed and applied to the study of $^{98\mathrm{m}}\mathrm{Mo}$ and $^{98\mathrm{m}}\mathrm{Zr}$.\\
Further experiments using direct $\gamma$-ray spectroscopy of the $2\gamma$ decay branch in $^{72}\mathrm{Ge}$ should elucidate the role of the magnetic and electric-dipole  terms. In view of the shorter than expected lifetime, leading to a larger branching ratio for the $2\gamma$ decay channel in neutral atoms, such an experiment seems nowadays feasible utilizing high-efficiency gamma-ray spectrometers.

\begin{acknowledgments}
The authors thank the GSI accelerator team for providing excellent technical support. The authors also thank G.\,Hudson-Chang for proofreading. B.\,A.\,B. acknowledges the support from the NSF grant PHY-2110365. P.\,K. acknowledges the support from BMBF under grant NuSTAR.DA 05P19RDFN1. T.\,Y. was partly supported by JSPS KAKENHI Grant Number T23KK0055. Work at ANL is supported by the U.S. Department of Energy, Office of Science, Office of Nuclear Physics, under contract No. DE-AC02-06CH11357. This work was supported by the Slovenian Research and Innovation Agency under Grants No. I0-E005 and No. P1-0102, by the European Research Council (ERC) under the European Union’s Horizon 2020 research and innovation programme (ERC-AdG NECTAR, grant agreement No 884715; ERC-CoG ASTRUm, grant agreement No 68284), the State of Hesse (Germany) within the Research Cluster ELEMENTS (Project ID 500/10.006), and by the STFC (UK).
\end{acknowledgments}

\end{document}